\documentclass[submitting]{nst}

\usepackage{subfigure,dcolumn}
\usepackage[T2A,T1]{fontenc}
\usepackage[russian,english]{babel}
\usepackage{todonotes}

\usepackage{listings}
\begin{document}

\title{Simulation study of BESIII with stitched CMOS pixel detector using \textsc{acts}}\thanks{Supported by the National Natural Science Foundation of China (Nos.~U2032203, 12275296, 12275297, 12075142, 12175256, 12035009) and National Key R\&D Program of China (No.~2020YFA0406302)}

\author{Yi Liu}
\affiliation{School of Physics and Microelectronics, Zhengzhou University, Zhengzhou, Henan, 450001, China}
\author{Xiaocong Ai}
\email[Corresponding author, ]{xiaocongai@zzu.edu.cn}
\affiliation{School of Physics and Microelectronics, Zhengzhou University, Zhengzhou, Henan, 450001, China}
\author{Guangyan Xiao}
\affiliation{School of Physics, Nanjing University, Nanjing 210093, Jiangsu, China}
\author{Yaxuan Li}
\affiliation{School of Physics, Nankai University, Tianjin 300071, China}
\author{Linghui Wu}
\email[Corresponding author, ]{wulh@ihep.ac.cn}
\affiliation{Institute of High Energy Physics, Chinese Academy of Sciences, 19B Yuquan Road, Shijingshan District, Beijing, China}
\author{Liangliang Wang}
\affiliation{Institute of High Energy Physics, Chinese Academy of Sciences, 19B Yuquan Road, Shijingshan District, Beijing, China}
\author{Jianing Dong}
\affiliation{Research Center for Particle Science and Technology, Institute of Frontier and Interdisciplinary Science, Shandong University, Qingdao 266237, Shandong, China}
\author{Mingyi Dong}
\affiliation{Institute of High Energy Physics, Chinese Academy of Sciences, 19B Yuquan Road, Shijingshan District, Beijing, China}
\affiliation{University of Chinese Academy of Sciences, 19A Yuquan Road, Shijingshan District, Beijing, China}
\author{Qinglin Geng}
\affiliation{Research Center for Particle Science and Technology, Institute of Frontier and Interdisciplinary Science, Shandong University, Qingdao 266237, Shandong, China}
\author{Min Luo}
\affiliation{School of Information Science \& Engineering, Harbin Institute of Technology, Weihai 264209, China}
\author{Yan Niu}
\affiliation{Research Center for Particle Science and Technology, Institute of Frontier and Interdisciplinary Science, Shandong University, Qingdao 266237, Shandong, China}
\author{Anqing Wang}
\affiliation{Research Center for Particle Science and Technology, Institute of Frontier and Interdisciplinary Science, Shandong University, Qingdao 266237, Shandong, China}
\author{Chenxu Wang}
\affiliation{School of Information Science \& Engineering, Harbin Institute of Technology, Weihai 264209, China}
\author{Meng Wang}
\affiliation{Research Center for Particle Science and Technology, Institute of Frontier and Interdisciplinary Science, Shandong University, Qingdao 266237, Shandong, China}

\author{Lei Zhang}
\affiliation{School of Physics, Nanjing University, Nanjing 210093, Jiangsu, China}
\author{Liang Zhang}
\affiliation{Research Center for Particle Science and Technology, Institute of Frontier and Interdisciplinary Science, Shandong University, Qingdao 266237, Shandong, China}
\author{Ruikai Zhang}
\affiliation{School of Information Science \& Engineering, Harbin Institute of Technology, Weihai 264209, China}
\author{Yao Zhang}
\affiliation{Institute of High Energy Physics, Chinese Academy of Sciences, 19B Yuquan Road, Shijingshan District, Beijing, China}
\author{Minggang Zhao}
\affiliation{School of Physics, Nankai University, Tianjin 300071, China}
\author{Yang Zhou}
\affiliation{Institute of High Energy Physics, Chinese Academy of Sciences, 19B Yuquan Road, Shijingshan District, Beijing, China}

\begin{abstract}
 Reconstruction of tracks of charged particles with high precision is very crucial for HEP experiments to achieve their physics goals. As the tracking detector of BESIII experiment, the BESIII drift chamber has suffered from aging effects resulting in degraded tracking performance after operation for about 15 years. To preserve and enhance the tracking performance of BESIII, one of the proposals is to add one layer of thin CMOS pixel sensor in cylindrical shape based on the state-of-the-art stitching technology, between the beam pipe and the drift chamber. The improvement of tracking performance of BESIII with such an additional pixel detector compared to that with only the existing drift chamber is studied using the modern common tracking software \textsc{acts}, which provides a set of detector-agnostic and highly performant tracking algorithms that have demonstrated promising performance for a few high energy physics and nuclear physics experiments.
\end{abstract}


\maketitle

%%%%%%%%%%%%%%%%%%%%%%%%%%%%%%%%%%%%%%%%%%%%%%%%%%%%%%%%%%%%%%%%%%
\section{Introduction}
\label{sec:intro}

The Beijing Spectrometer (BESIII)~\cite{ABLIKIM2010345} at the Beijing Electron-Positron (BEPCII) Collider has been a great success producing prosperous physics results~\cite{Yuan2019} in the $\tau$-charm sector since 2009. However, after operation for about 15 years, the detectors at BESIII has been subjected to aging effects, which result in degraded performance of the detectors~\cite{Ablikim_2020}. 

The tracking detector at BESIII is the Multilayer Drift Chamber (MDC), which provides measurement of momentum and position of the charged tracks, and information of energy loss in unit path length, i.e.~d$E$/dx~\cite{Cao_2010}, of the charged tracks for particle identification. As shown in Ref.~\cite{Dong_2016,Ablikim_2020}, due to the beam-induced background with a hit rate up to 2 kHz/cm$^2$, the gain of the MDC cells in the first ten layers has shown an obvious decrease with a maximum decrease of about 39\% for the innermost layer cells in 2017, which further leads to reduction of the spatial and momentum resolution of the charged tracks. Since BESIII is not expected to complete its mission in foreseen years~\cite{Ablikim_2020,Yuan2019}, the track reconstruction performance must be preserved and enhanced in order to not compromise the physics goals of BESIII, along with possible upgrade of the tracking system with state-of-the-art technologies for detection of charged particles. To get well prepared for the potential malfunction of the MDC due to aforementioned aging problem, plans of upgrading the BESIII inner tracker based on different technologies have been proposed~\cite{Ablikim_2020}. This includes replacement of the inner tracker with a new inner drift chamber~\cite{Xie_2016} or a cylindrical gas electron multiplier (CGEM) tracker~\cite{BORTONE2023167957}, which has attractive features such as high counting rate tolerance and low sensitivity to aging. 

Compared to the gaseous detectors, the silicon pixel detector has excellent spatial resolution and good radiation resistance. Therefore, one of the additional options for the BESIII inner tracker upgrade is to use large-area thin Complementary Metal Oxide Semiconductor (CMOS) pixel sensor with good spatial resolution based on the cutting-edge stitching technology, which has already been used to produce CMOS pixel sensors for medical imaging applications~\cite{las_2009,dexela_2012,farrier_2009}. Recently, the studies of designing a first wafer-scale stitched sensor prototype, the MOSS (Monolithic Stitched Sensor) chip, towards an improved vertex detector, i.e.~the ITS3, at ALICE experiment, are presented in Ref.~\cite{AGLIERIRINELLA2023168018}.

A Common Tracking Software (\textsc{acts})~\cite{Ai2022} is a common High Energy Physics (HEP) software, providing a set of detector-agnostic, performant and modular tools for the track reconstruction in HEP using modern software technologies to facilitate concurrency, usability, maintenance and extendability to tackle the tracking challenges foreseen in HEP in future. So far, \textsc{acts} has been used for the track reconstruction at ATLAS~\cite{ATL-PHYS-PUB-2021-012}, FASER~\cite{PhysRevLett.131.031801}, sPHENIX~\cite{Osborn2021}, STCF~\cite{Ai_2023} etc., for different types of tracking detectors. In particular, promising tracking performance of \textsc{acts} for a tracking system with a drift chamber is firstly presented in Ref.~\cite{Ai_2023}.

In this study, the spatial and momentum resolution of the BESIII MDC with an additional one-layer stitched cylindrical CMOS pixel detector inserted between the beam pipe and the inner wall of the MDC is studied using \textsc{acts} as the tracking software. The performance is compared to that of the current BESIII tracking detector with only the MDC based on the BESIII Offline Software System (\textsc{boss})~\cite{BOSS}. The manuscript is organized as follows. In Section~\ref{sec:besiii}, the BESIII MDC and the proposed pixel detector based on the cylindrical CMOS pixel sensor are introduced. Section~\ref{sec:software} introduces the tracking strategies in \textsc{boss} and \textsc{acts}. Improvement of the spatial and momentum resolution at BESIII if the additional pixel detector is added is presented in Section~\ref{sec:perf}. A brief conclusion is provided in Section~\ref{sec:con}.

%%%%%%%%%%%%%%%%%%%%%%%%%%%%%%%%%%%%%%%%%%%%%%%%%%%%%%%%%%%%%%%%%%
\section{The BESIII tracking detector and stitched CMOS pixel detector}
\label{sec:besiii}

The BESIII MDC is a cylindrical chamber operating with a helium-based mixture gas (He/C$_3$H$_8$ = 60:40) and immersed in a 1 T magnetic field. The inner radius and outer radius of the MDC is 64 mm and 819 mm, respectively. The length of the wires ranges from 774 mm for the innermost layer to 2400 mm for the outermost layer. The drift cells which are almost square shaped are arranged in 43 circular layers, which alternate between the stereo layers and axial layers, i.e.~in the order of 8 stereo
layers, 12 axial layers, 16 stereo layers and 7 axial
layers. The cell dimension is about 12~mm $\times$ 12~mm for the 8 inner layers, which compose the inner chamber, and 16.2~mm $\times$ 16.2~mm for the 35 outer layers, which compose the outer chamber. 

The CMOS pixel sensor has been widely used for vertex detectors in HEP, due to its excellent spatial resolution down to a few $\mu$m, tolerance of high hit rate up to $10^8$ Hz/cm$^2$, and good detection efficiency and radiation resistance. For example, the CMOS pixel sensor has been used for the inner tracker or vertex detector of the STAR experiment~\cite{DOROKHOV2011174}, the ALICE experiment~\cite{AGLIERIRINELLA2017583}, and the CEPC project~\cite{DONG2023167967}. Despite all the advantages of the CMOS pixel sensor, the tracking resolution of a traditional CMOS pixel detector, in particular for low momentum tracks, can be limited by the irreducible material budget of the support structure and cooling pipe, as shown in Ref.~\cite{DONG2019287}, where the MAPS-based CMOS pixel sensor is proposed for the upgrade of BESIII inner chamber. Therefore, tracking based on traditional CMOS pixel sensors can be challenging at BESIII, where tracks with transverse momentum down to 150 MeV are required to be reconstructed with good resolution.

In recent years, the innovative stitching technology emerged with the aim of producing large-area sensor in one wafer, which can be further thinned to about 50 $\mu$m. With such thickness, the wafer can be curved into cylindrical shape hence vastly simplifying the support structure and reducing the material budget.
Based on such stitching technology, it's possible to construct a vertex detector using large-area thin CMOS pixel sensors with cylindrical shape and simplified support structure. In this study, a vertex detector which consists of one layer of stitched cylindrical pixel sensor with thickness of 50 $\mu$m, resolution of 8.66 $\mu$m in the $r$-$\phi$ direction and 57.74 $\mu$m in the $z$ direction, inserted between the beam pipe and MDC, is considered and denoted as the "Pixel" detector.

\begin{figure}[!htb]
\includegraphics
  [width=\hsize]
  {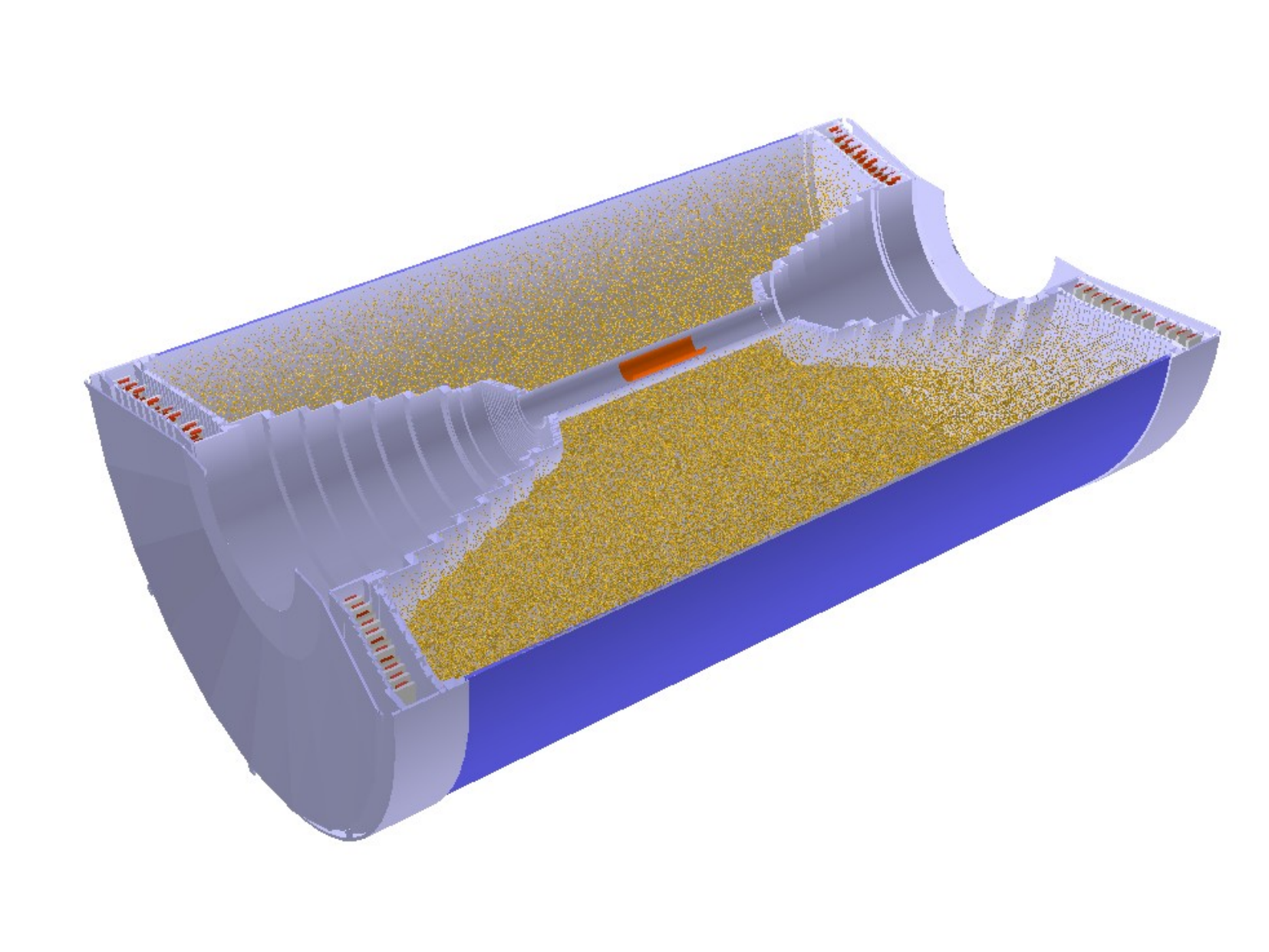}
\caption{A half-clipped view of geometry of the BESIII MDC (cell wires are shown in golden colors) and the Pixel detector (shown in orange color). The beam pipe is not shown here.}
\label{fig:besiii_geom}
\end{figure}

The geometry of the BESIII MDC and the Pixel detector is shown in Figure~\ref{fig:besiii_geom}.

%%%%%%%%%%%%%%%%%%%%%%%%%%%%%%%%%%%%%%%%%%%%%%%%%%%%%%%%%
\section{Tracking with BESIII offline software and \textsc{acts}}
\label{sec:software}
The simulation study is based on samples generated using \textsc{boss}, which is the offline software framework for event generation, simulation, reconstruction, performance validation, physics analysis and visualization for the BESIII experiment. 

In \textsc{boss}, the decays of particles are modelled with \mbox{\textsc{evtgen}}~\cite{evtgen}, which can also be used to generate single particle samples. The detector description is based on the Geometry Description Markup Language (\textsc{gdml})~\cite{gdml} and the interaction of particles with detectors is simulated with \textsc{geant4}~\cite{AGOSTINELLI2003250}. 

The performance of the BESIII MDC with the additional Pixel detector is studied using the \textsc{acts} software, and compared to the performance of the BESIII MDC studied using the BESIII tracking software within \textsc{boss}.

\subsection{Tracking in \textsc{boss}}

The trajectory of a charged particle in magnetic field is parameterized using the helix track parameters at BESIII, as described in Ref.~\cite{Liu_2008}. Track finding and track fitting are two tasks in \textsc{boss} track reconstruction. Track finding is a pattern recognition problem of classifying measurements into subsets and creating track candidates, while track fitting is an estimation of the helix parameters. Two basic track finding algorithms are implemented in \textsc{boss}: the template matching algorithm (\textsc{pat})~\cite{PAT} and track segment finder algorithm (\textsc{tsf})~\cite{Liu_2008}. Both a specialized track finding method called TCurlFinder~\cite{Jia_2010} and the global track finding method based on Hough transform (\textsc{hough})~\cite{HOUGH} have been implemented to salvage low transverse momentum tracks with $p_T <$ 0.12 GeV. Besides, an extended segment construction scheme which can achieve higher efficiency for low transverse momentum tracks has been developed~\cite{Ma_2013}. The found tracks using these algorithms are combined and fed into the Kalman Filter~\cite{Wang_2009} algorithm for track fitting. Further track extrapolation is performed to obtained the track parameters at other subdetectors at BESIII.

\subsection{Tracking with \textsc{acts}}

A detailed introduction to \textsc{acts}, including geometry description, parameterization of track parameters and measurements, and tracking algorithms etc., can be found in Ref.~\cite{Ai2022}. The implementation of \textsc{acts} for track reconstruction for the BESIII tracking detector is outlined below.

To use \textsc{acts} for track reconstruction, the detector geometry with detailed description of detector material and placement, e.g.~the \textsc{geant4}-based detector geometry, needs to be transformed into \textsc{acts} internal geometry, i.e.~\textsc{acts} tracking geometry, which has a simplified description of the passive detector material to facilitate fast navigation and track reconstruction. The \textsc{tgeo}~\cite{BRUN2003676} plugin in \textsc{acts} is used to transform the \textsc{tgeo} version of the BESIII detector geometry, which is created based on \textsc{gdml} exported from the \textsc{geant4}-based geometry, into \textsc{acts} tracking geometry with material mapped to the auxiliary surfaces. The one pixel layer of the Pixel detector is converted to a cylinder surface in \textsc{acts} and the 43 layers of signal wires of the MDC are converted to 43 layers of line surfaces in \textsc{acts}.

The hits on the MDC after digitization are transformed into one-dimensional measurements associated to the line surfaces in \textsc{acts} describing the drift distance of a drift chamber. The hits on the Pixel after simulation are transformed into two-dimensional measurements by smearing the simulated hits with the resolution of the pixel detector using Gaussian functions. Those measurements are associated to the \textsc{acts} cylinder surfaces, where the local $x$ ($y$) coordinate represents the $r\cdot\varphi$ ($z$) in the cylinder frame, i.e.~$r$ represents the radius of the cylinder, $\varphi$ represents the azimuthal angle of the position on the cylinder and $z$ is the coordinate in the $z$ direction.

The BESIII magnetic field is transformed into \textsc{acts} interpolated magnetic field using the field map of BESIII magnetic field. With an interpolated field provider in \textsc{acts}, the value of the magnetic field for any given position is calculated by interpolating from a grid of known values, e.g.~eight corner points of a field cell in three-dimensional coordinate system.

The track fitting is performed using the combinatorial Kalman Filter~\cite{Fruhwirth2021} algorithm in \textsc{acts}, which has the capability of rejecting noise hits based on the $\chi^2$ calculated using the distance between the hit and the predicted track parameters, and the covariance of the hit and the predicted track parameters. If multiple hits are found to be compatible with the predicted track parameters, the hit with the best $\chi^2$ is used to filter the track parameters. The smoothed track parameters at the first measurement are eventually extrapolated to the beam line to obtain the estimated track parameters at the interaction point.

%%%%%%%%%%%%%%%%%%%%%%%%%%%%%%%%%%%%%%%%%%%%%%%%%%%%%%%%%%%%%%%%%%

\section{Track reconstruction performance}
\label{sec:perf}

\subsection{Monte-Carlo sample generation}
The impact of the Pixel detector on the resolution of the track parameters is studied using single $\mu^-$ and single $\pi^-$ samples. The samples are generated with fixed transverse momentum $p_T$, cos$\theta$ ($\theta$ is the polar angle) uniformly distributed between [-0.8, 0.8] , and azimuthal angle $\phi$ uniformly distributed in the range of [0, $2\pi$]. The random background hits in the MDC due to beam related background or electronic noises are generated as in the standard simulation sample production at BESIII. Since dedicated background noise model for the Pixel detector is not available yet, no background noise is considered for the inserted pixel layer in this study.

\subsection{Track parameter resolution}
\begin{figure*}[!htb]
\includegraphics
  [width=0.33\hsize]
  {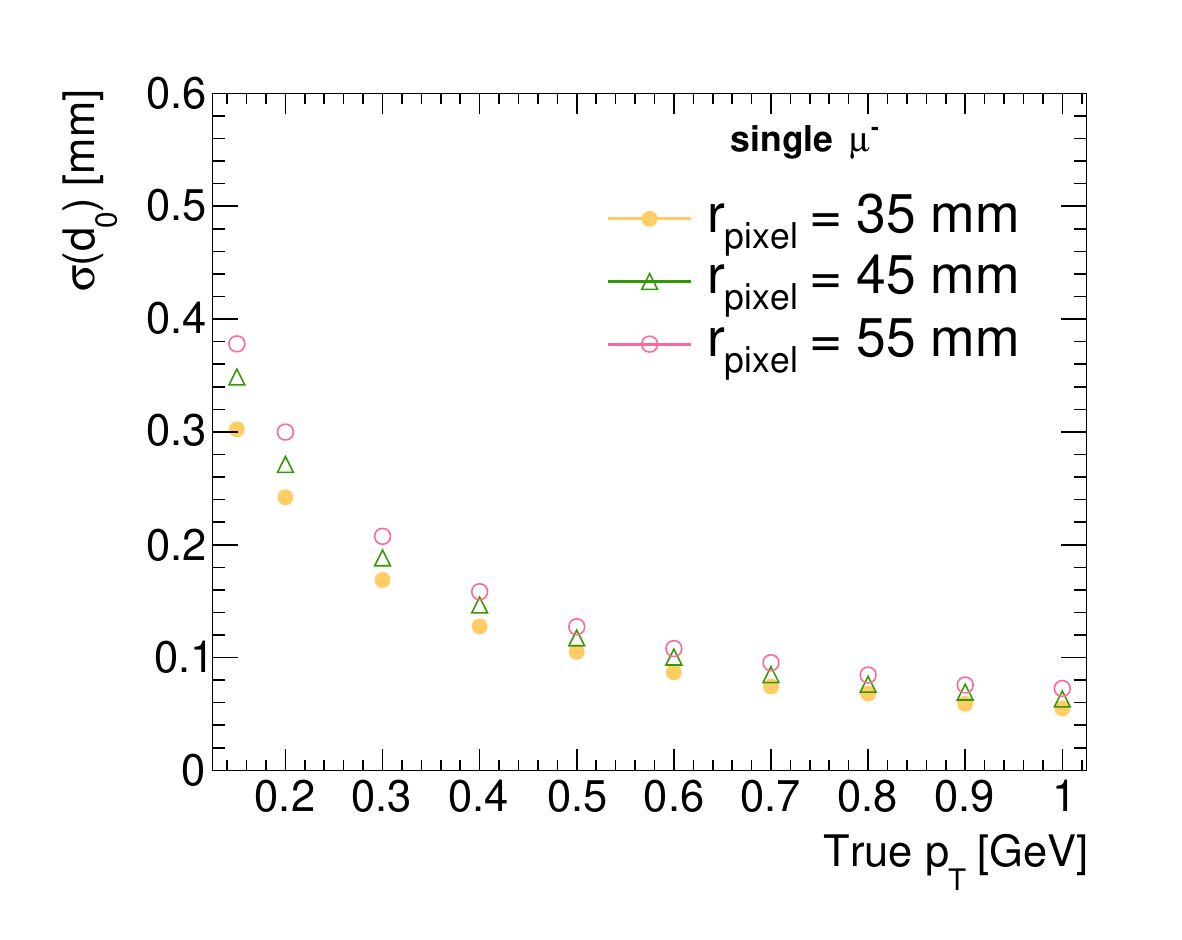}
\includegraphics
   [width=0.33\hsize]
  {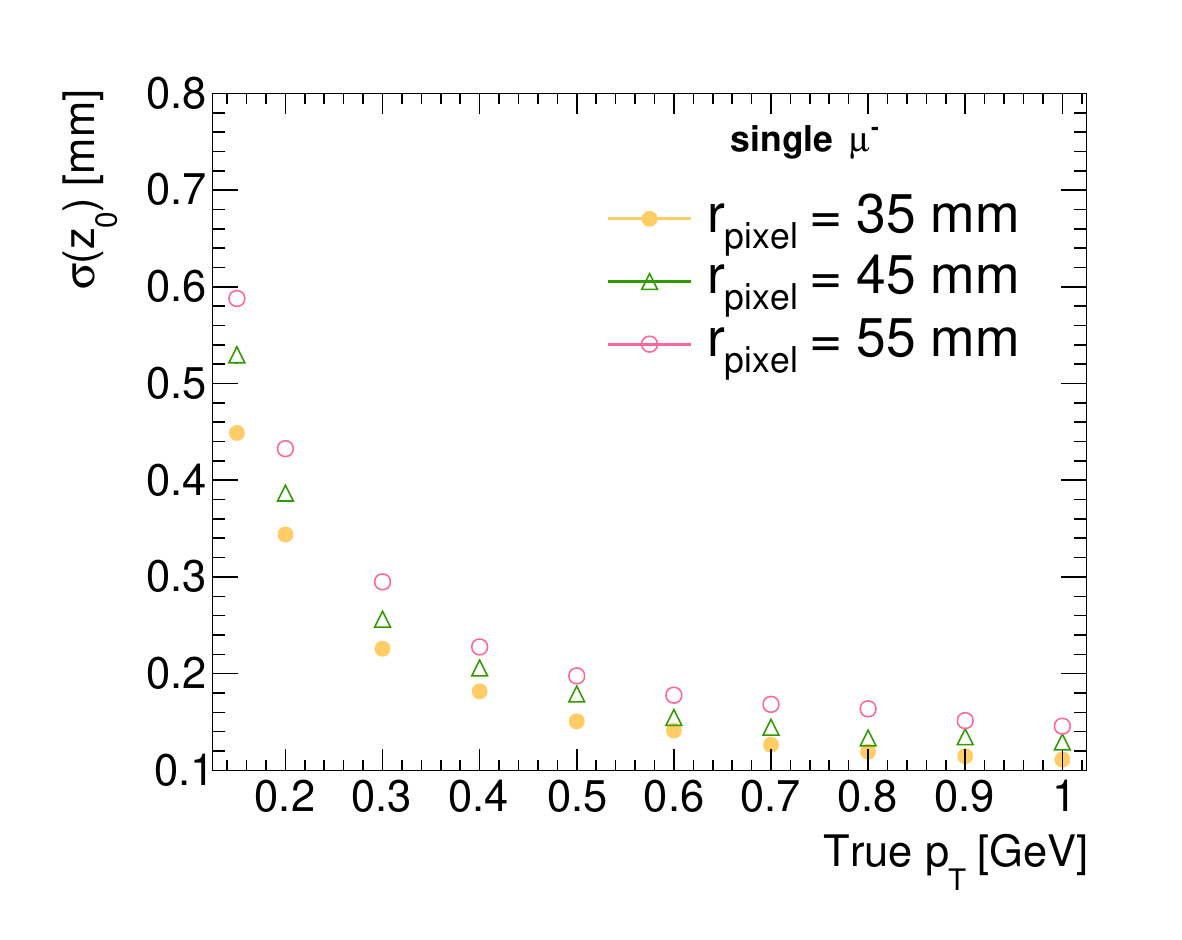}
\includegraphics
   [width=0.33\hsize]
  {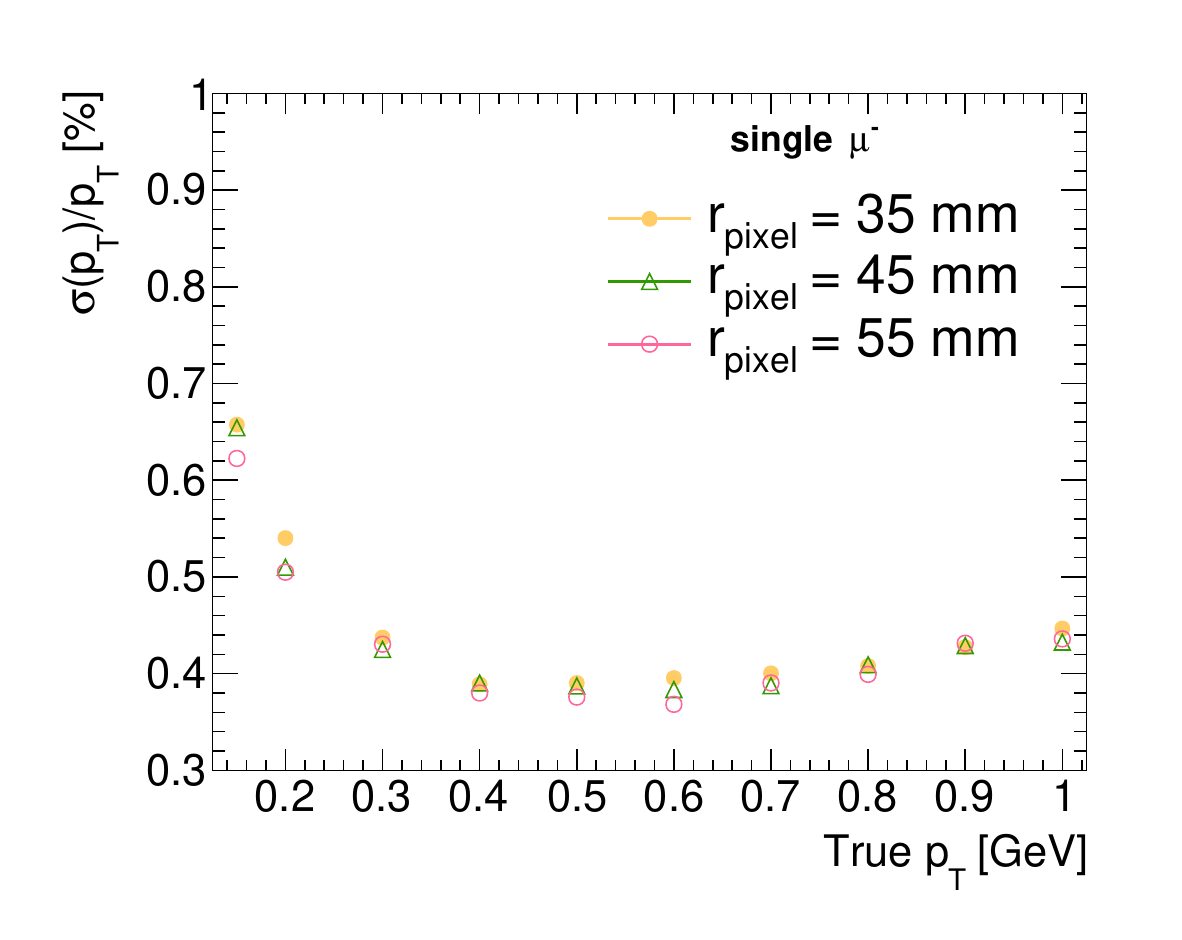} \\
  
\includegraphics
    [width=0.33\hsize]
  {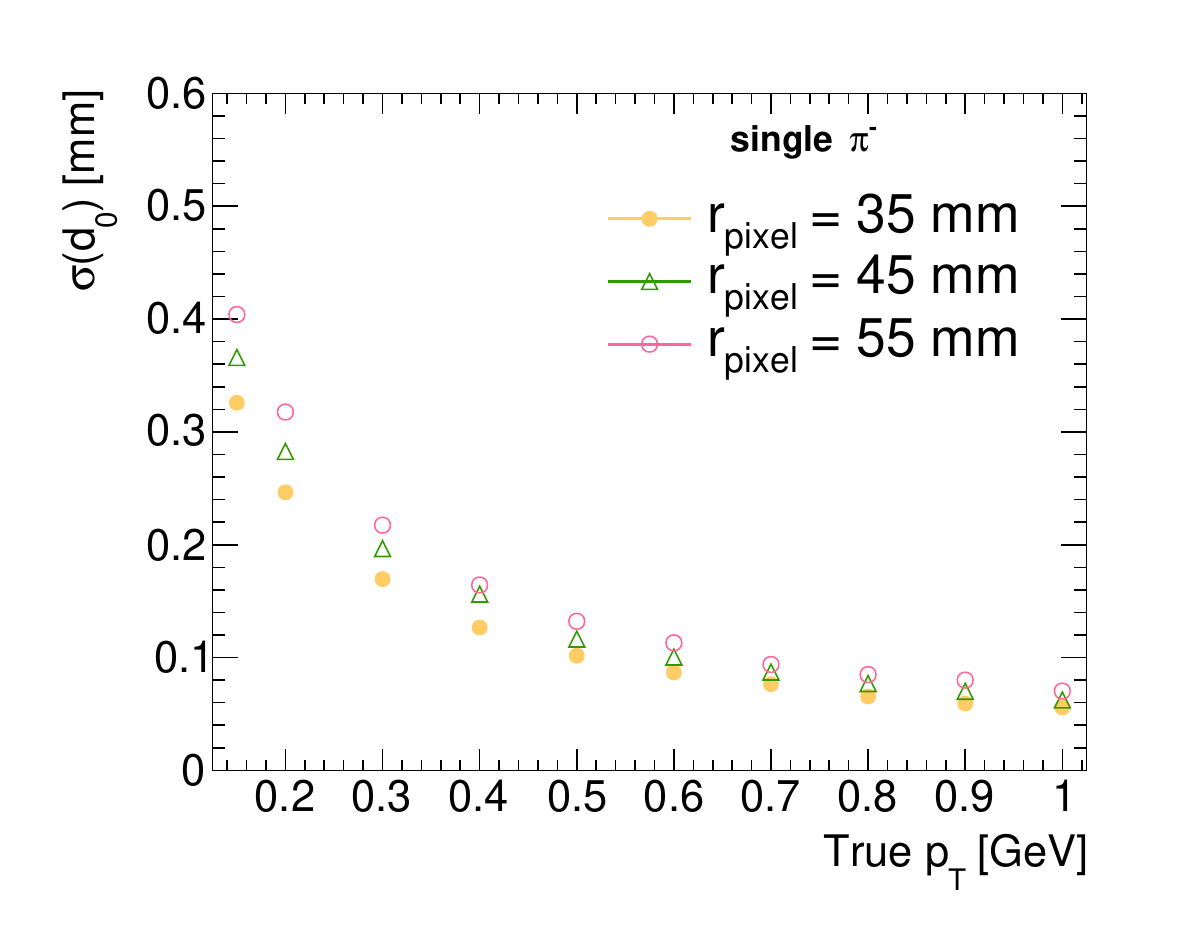}
\includegraphics
   [width=0.33\hsize]
  {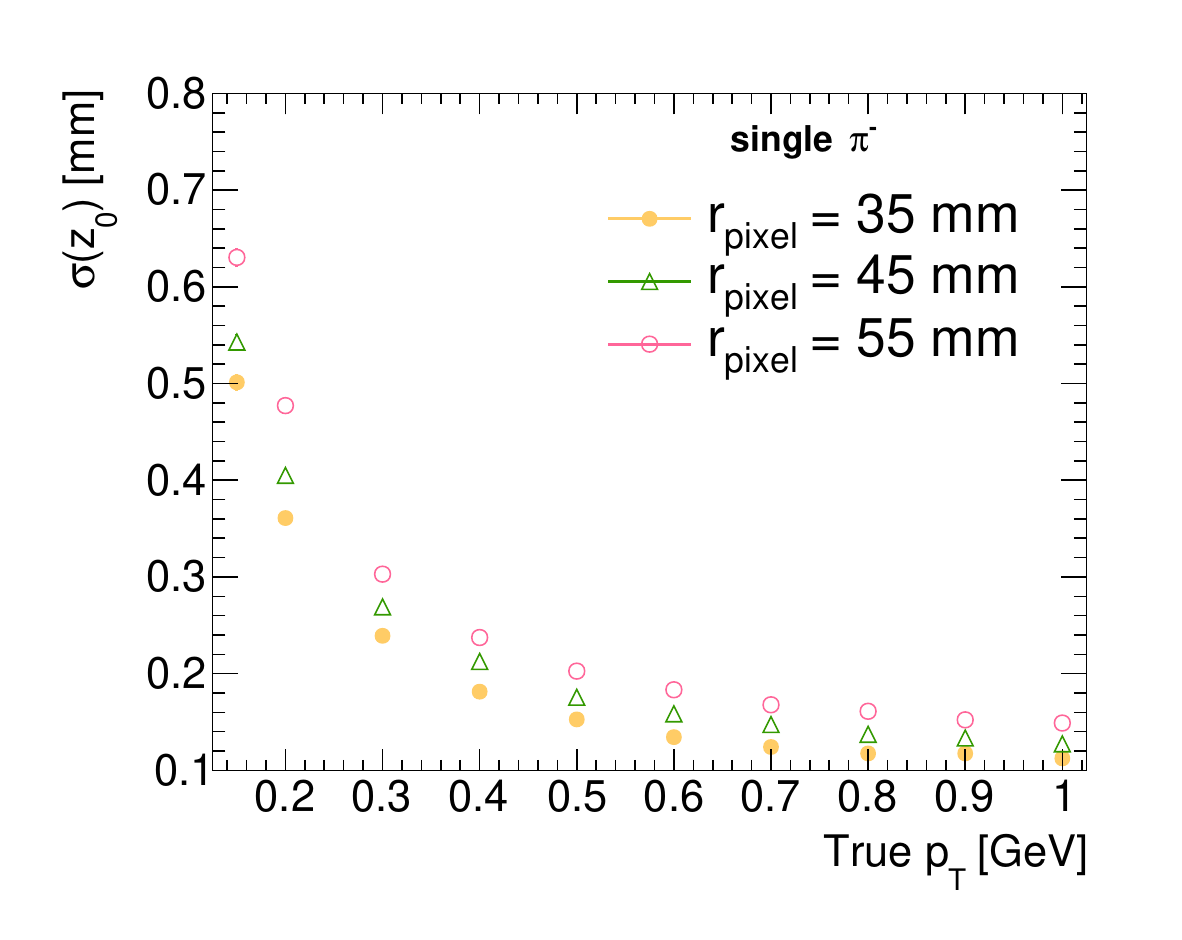}
\includegraphics
   [width=0.33\hsize]
  {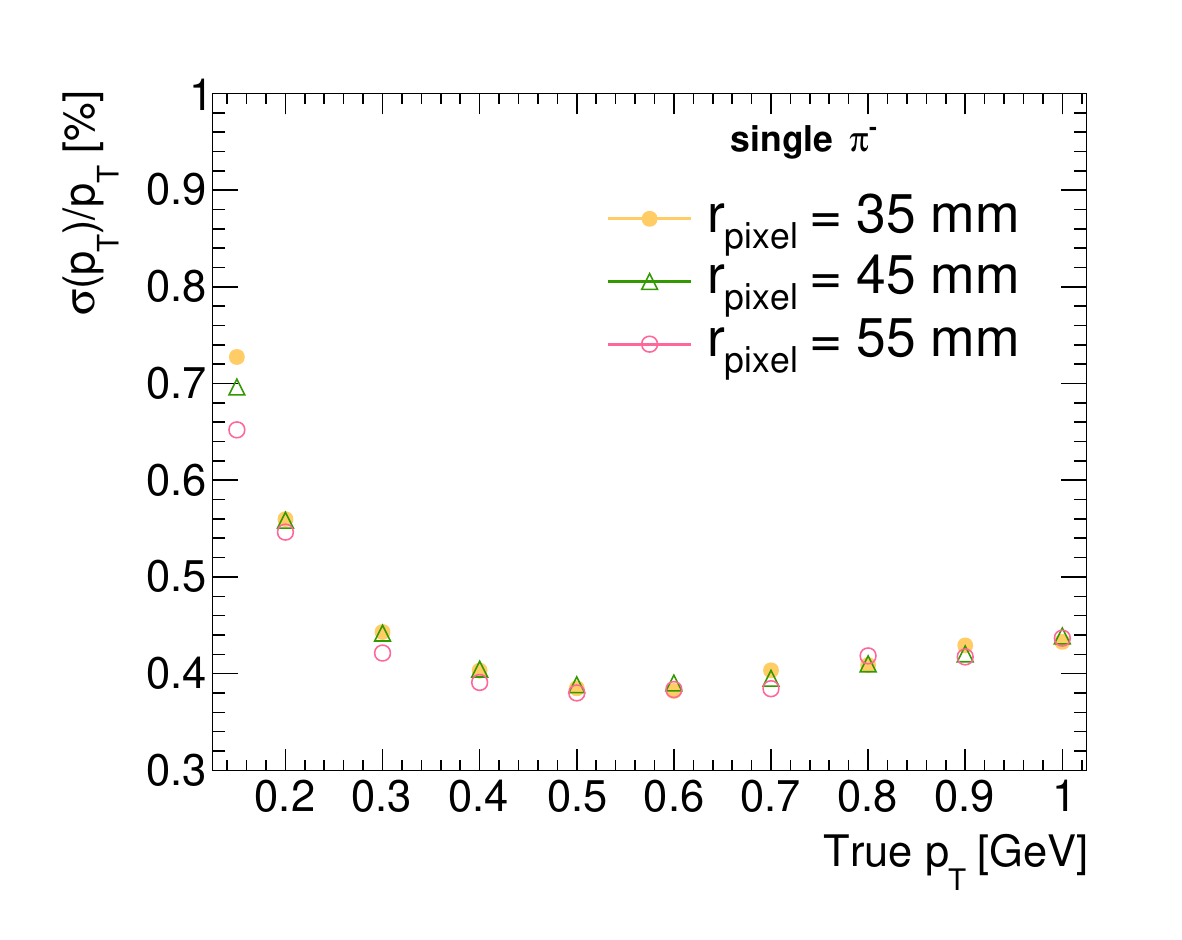} 
\caption{The resolution of $d_0$ (left panels), $z_0$ (middle panels) and relative resolution of $p_T$ (right panels) for single $\mu^-$ (top panels) and single $\pi^-$ (bottom panels) as a function of particle $p_T$ for the BESIII MDC with an additional pixel layer placed with $r_{\textrm{pixel}}$ = 35 mm (yellow dot), 45 mm (green triangle) and 55 mm (pink circle), respectively.}
\label{fig:track_reso_pixelR}
\end{figure*}

The resolution of the impact track parameters, $d_0$, $d_z$ and the relative resolution of the transverse momentum $p_T$ as a function of $p_T$ for the BESIII MDC with an additional pixel layer, which is placed with the radius $r_{\textrm{pixel}}$ at three different values, i.e.~35 mm, 45 mm or 55 mm, is shown in Figure~\ref{fig:track_reso_pixelR}. While $r_{\textrm{pixel}}$ has little impact on the resolution of $p_T$, the spatial resolution is better with smaller $r_{\textrm{pixel}}$. With $r_{\textrm{pixel}}$ = 35 mm, the resolution of $d_0$ and $d_z$ is 60 $\mu$m and 120 $\mu$m for $\mu^-$ and $\pi^-$ with $p_T$ = 1 GeV, respectively, and the relative resolution of 44\% for $p_T$ with $p_T$ = 1 GeV is achieved.

\begin{figure*}[!htb]
\includegraphics
  [width=0.33\hsize]
  {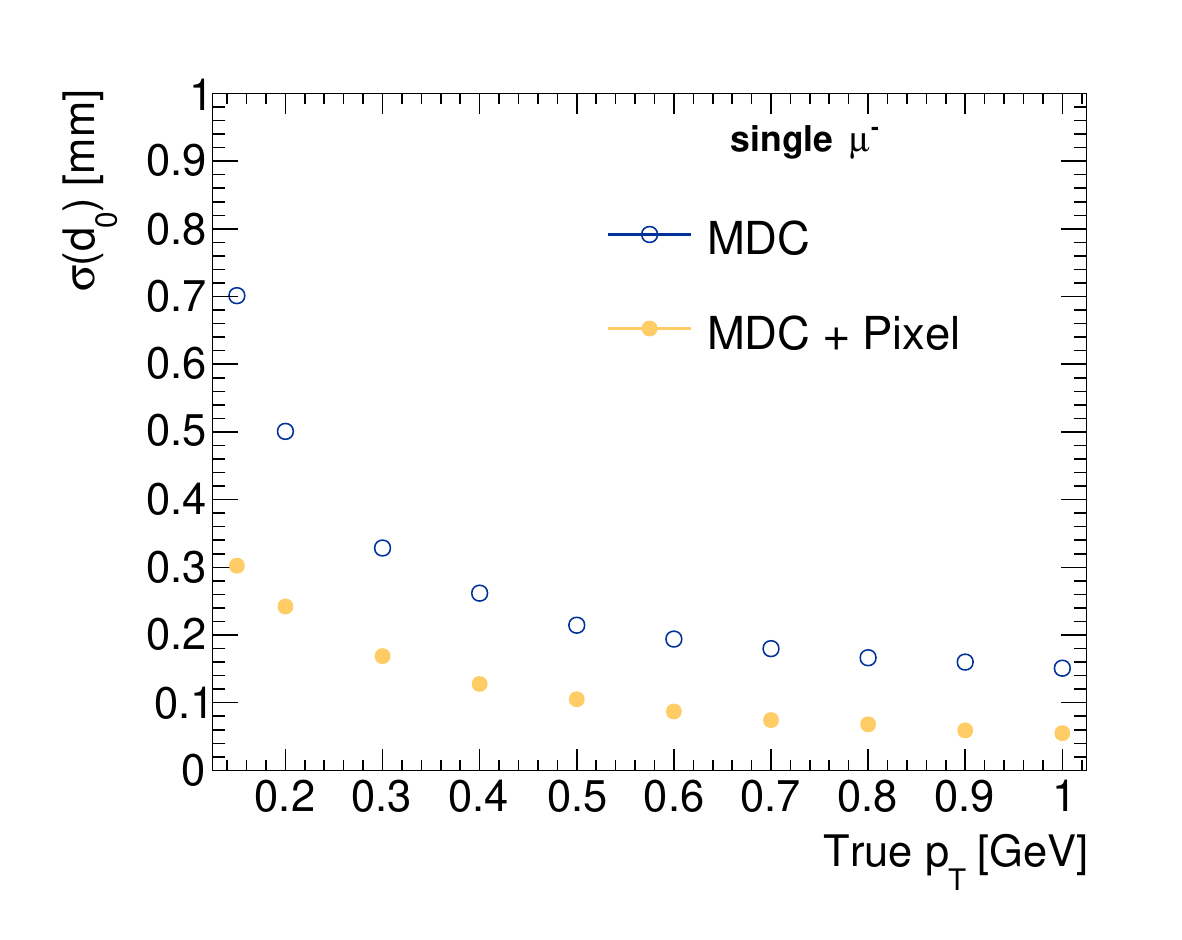}
\includegraphics
   [width=0.33\hsize]
  {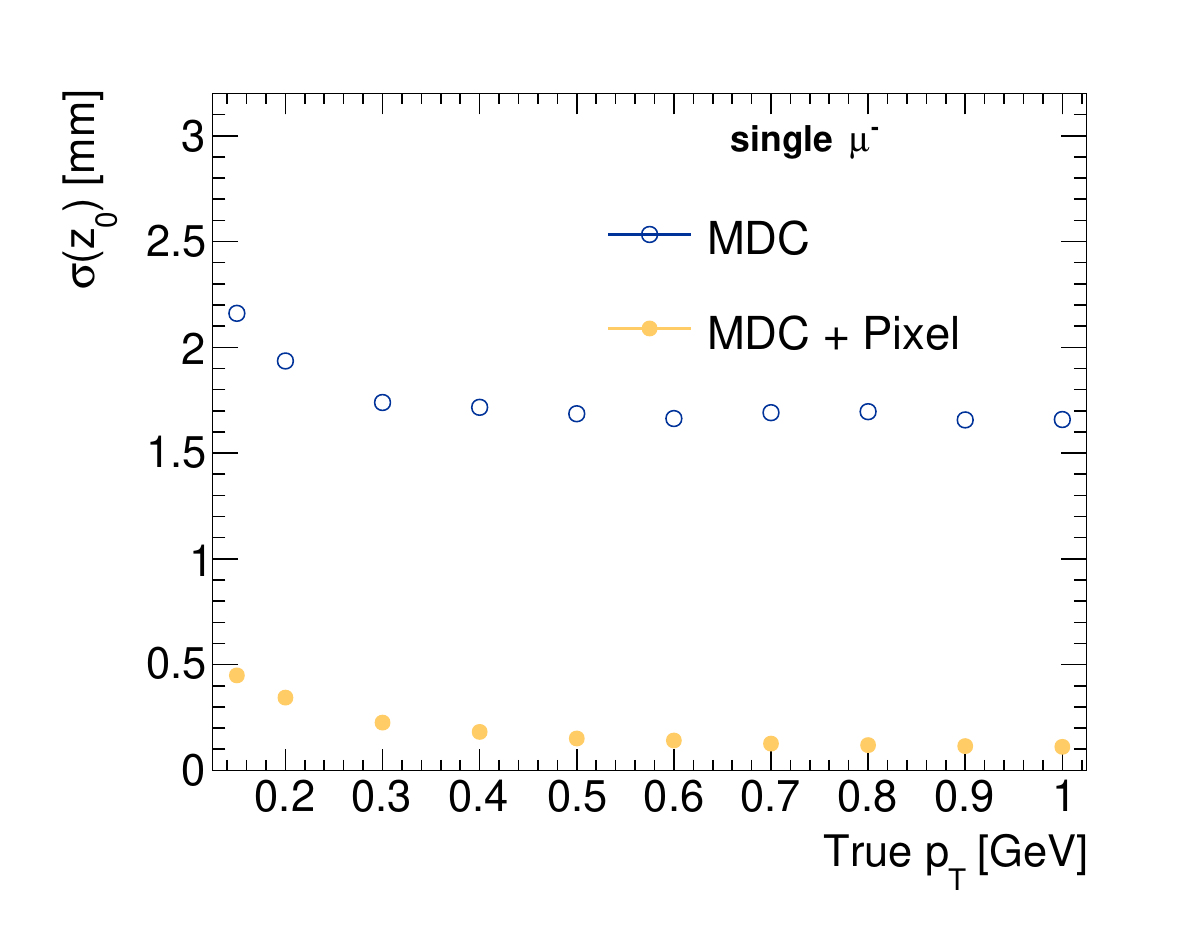}
\includegraphics
   [width=0.33\hsize]
  {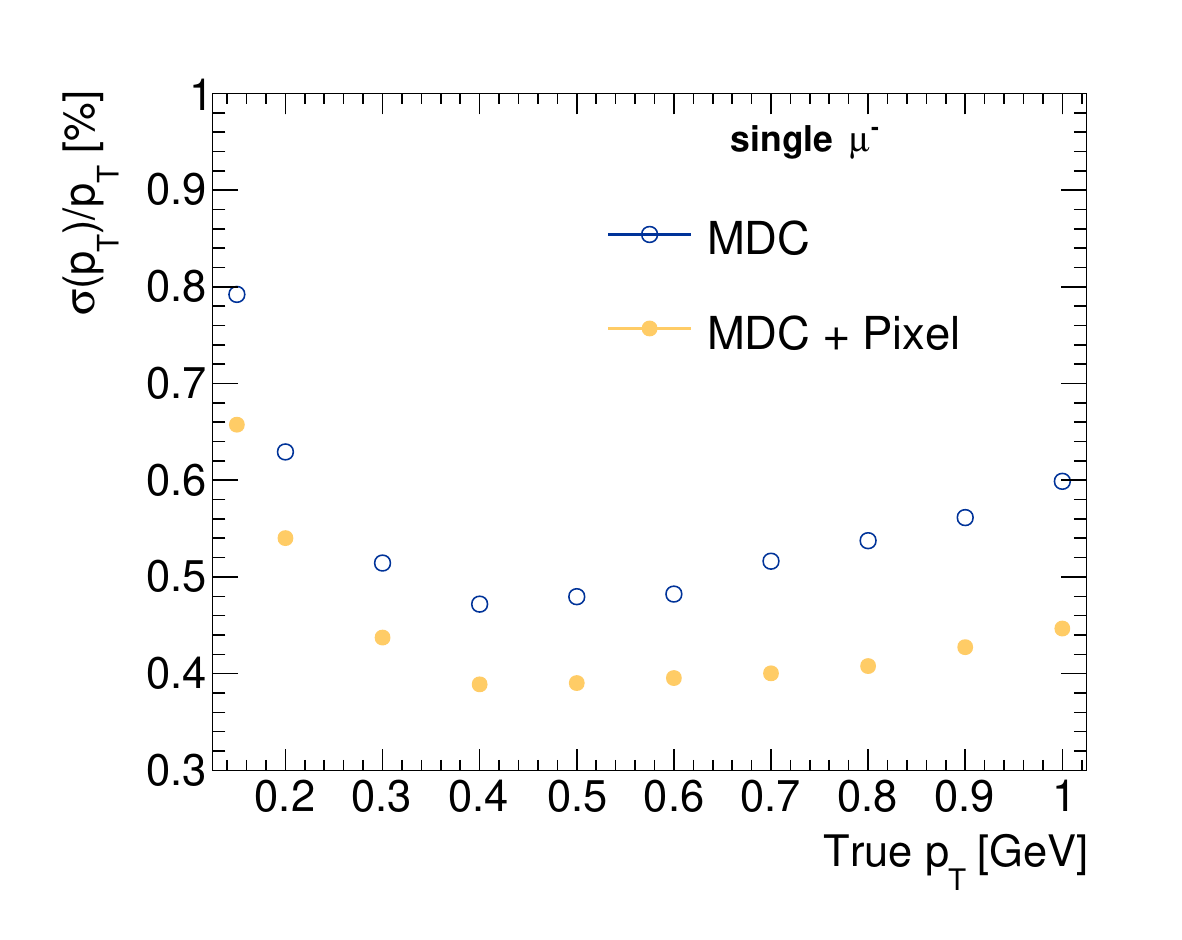} \\
  
\includegraphics
    [width=0.33\hsize]
  {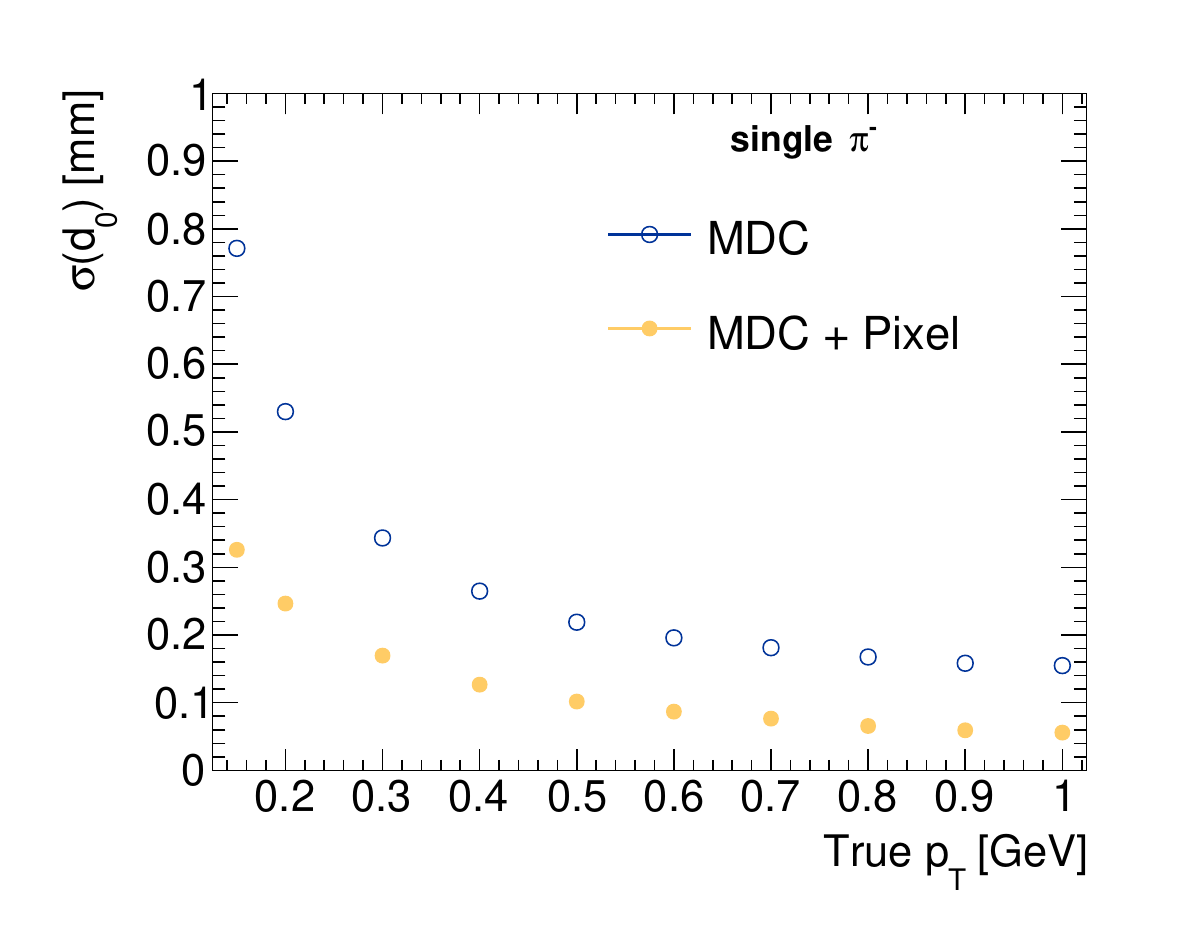}
\includegraphics
   [width=0.33\hsize]
  {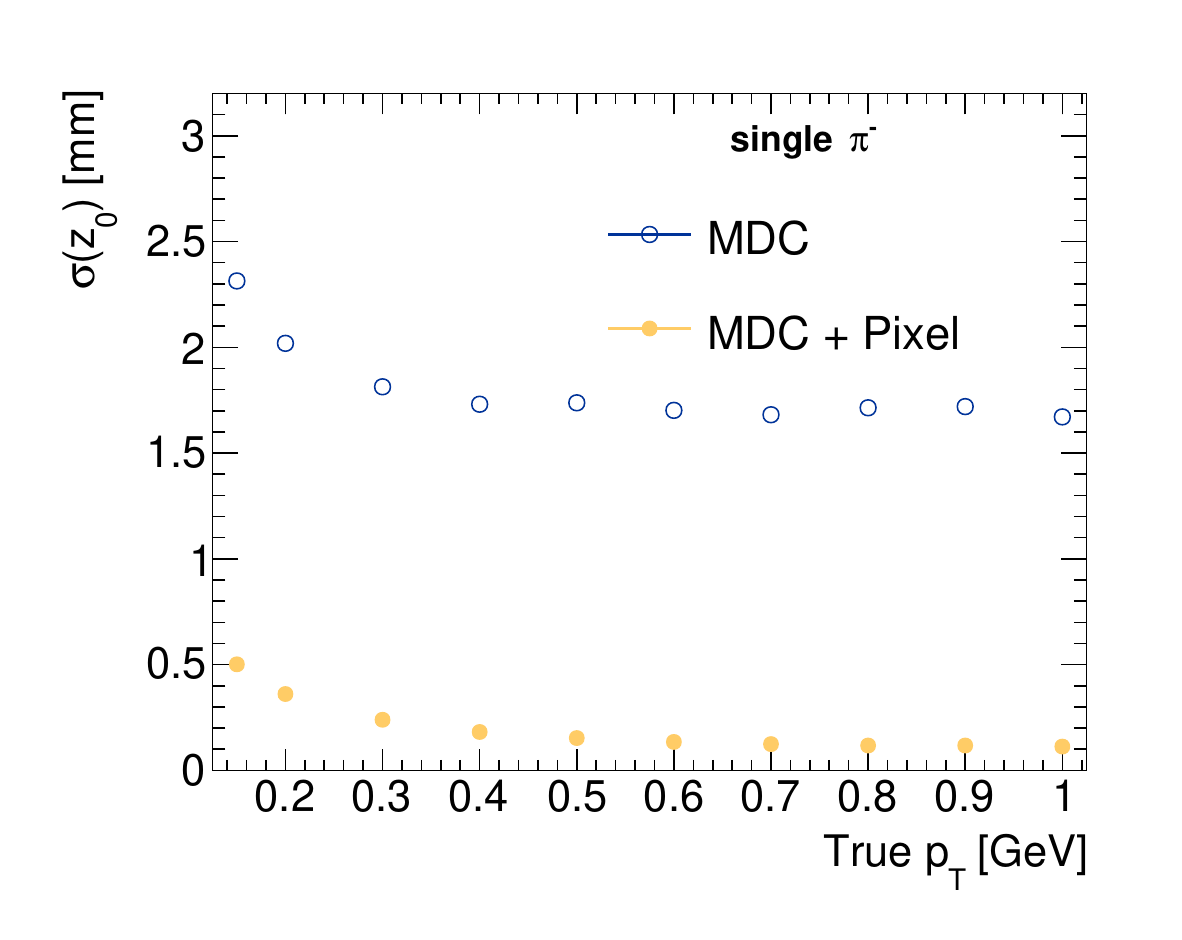}
\includegraphics
   [width=0.33\hsize]
  {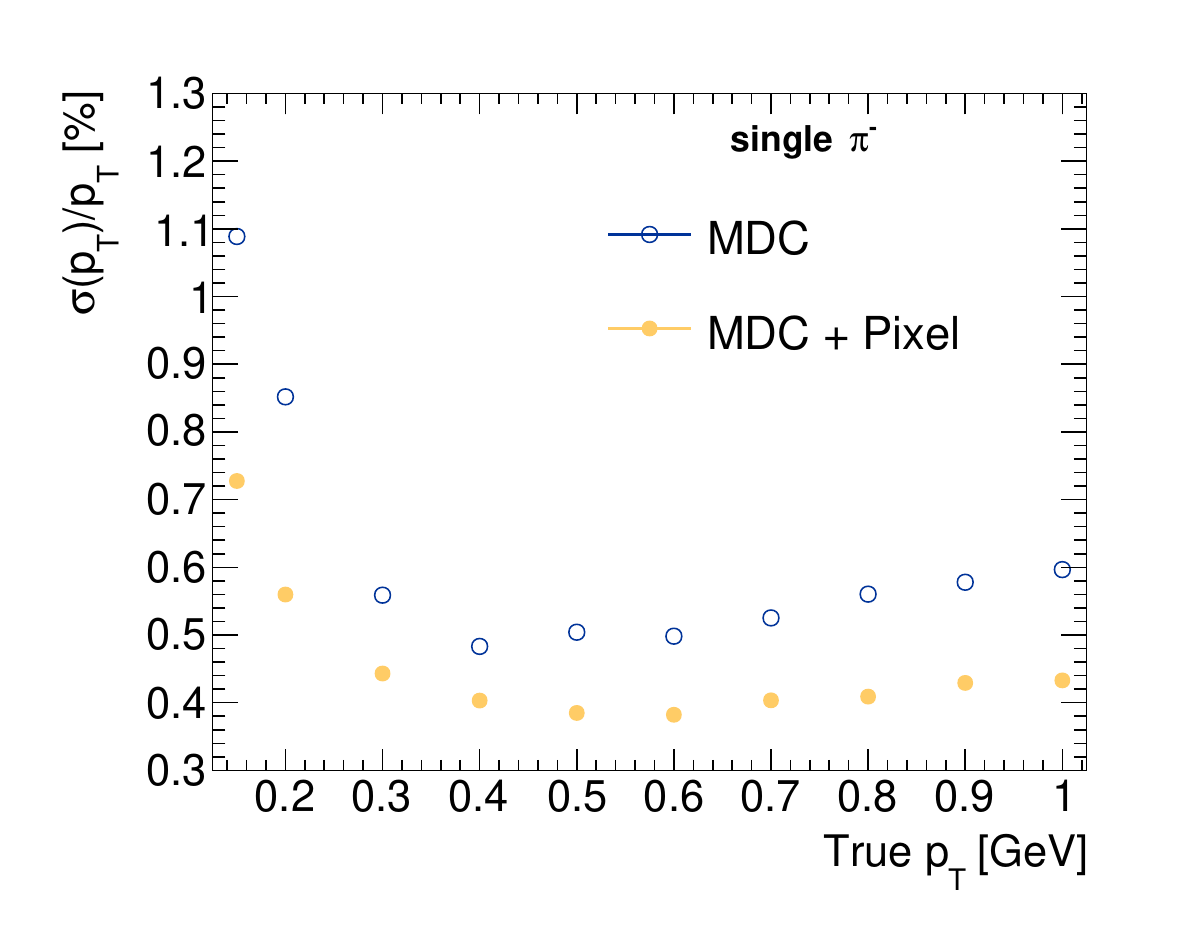} 
\caption{The resolution of $d_0$ (left panels), $z_0$ (middle panels) and relative resolution of $p_T$ (right panels) for single $\mu^-$ (top panels) and single $\pi^-$ (bottom panels) as a function of particle $p_T$. The blue circles and yellow dots represent the results with BESIII MDC and BESIII MDC with an additional pixel layer ($r_{\textrm{pixel}}$ = 35 mm), respectively.}
\label{fig:track_reso_comp}
\end{figure*}

A comparison of the resolution of $d_0$, $d_z$ and the relative resolution of $p_T$ between BESIII MDC and BESIII MDC with the pixel layer placed at $r_{\textrm{pixel}}$ = 35 mm as a function of $p_T$ , is shown in Figure~\ref{fig:track_reso_comp}. The resolution of $d_0$, $d_z$, $p_T$ can be improved by up to 67\%, 93\% and 32\%, respectively, if the proposed Pixel detector is added to the current BESIII tracking detector. 

\section{Conclusion}
\label{sec:con}

After operation for about 15 years, the tracking detector of BESIII experiment has suffered from aging effects and needs to be upgraded to preserve its tracking performance for BESIII to fullfill its remaining physics goals in the next few years. The possibility of inserting a one-layer pixel detector using the large-area thin CMOS pixel sensor in cylindrical shape based on the stitching technology, between the beam pipe and the MDC, was considered. The spatial and momentum resolution was studied using \textsc{acts} and compared to the performance of the current BESIII tracking detector obtained using the BESIII offline software. The proposal based on the stitched cylindrical CMOS pixel sensor was found to be very promising, i.e.~the resolution of $d_0$, $d_z$, $p_T$ can be improved by up to 67\%, 93\% and 32\%, respectively, with the additional pixel layer placed at an radius of 35 mm. Meanwhile, it's the first time that \textsc{acts} was used for track reconstruction for a drift chamber with calibrated measurements and realistic noise hits. In future studies, the noise model of this \mbox{state-of-the-art} pixel detector will be studied and the tracking performance taking into account the noise on the pixel detector will be investigated.

%%%%%%%%%%%%%%%%%%%%%%%%%%%%%%%%%%%%%%%%%%%%%%%%%%%%%%%%%%%%%%%%%%

\acknowledgments

This work is supported in part by National Natural Science Foundation
of China (NSFC) under Contracts Nos. U2032203, 12275296, 12275297, 12075142, 12175256, 12035009 and National Key R\&D Program of China under Grants No. 2020YFA0406302.

%\section{Bibliography}

\bibliography{bibfile}

%\begin{thebibliography}{99}
%\end{thebibliography}

\end{document}